\documentclass{aastex}
\usepackage{emulateapj5,apjfonts}

\newcommand{\pcmsq}{\mbox{cm$^{-2}$}}
\newcommand{\cmsq}{\mbox{cm$^{2}$}}
\newcommand{\ergsec}{\mbox{erg s$^{-1}$}}
\newcommand{\ergcms}{\mbox{erg cm$^{-2}$ s$^{-1}$}}

\newcommand{\Lx}{\mbox{$L_{\rm x}$}}

\newcommand{\nh}{\mbox{N$_H$}}
\newcommand{\chandra}{\textit{Chandra}}
\newcommand{\cxo}{\textit{Chandra X-ray Observatory}}
\newcommand{\rosat}{\textit{ROSAT}}
\newcommand{\xspec}{\textsc{XSPEC}}
\newcommand{\tool}{\it}
\newcommand{\xsoft}{\mbox{X$_{\rm soft}$}}
\newcommand{\xmed}{\mbox{X$_{\rm med}$}}
\newcommand{\xhard}{\mbox{X$_{\rm hard}$}}

\newcommand{\err}[2]{\small \ensuremath{^{+#1}_{-#2}}}
\newcommand{\ee}[2]{\ensuremath{#1\times 10^{#2}}}

\shorttitle{{\it Chandra} Observation of NGC 6440}
\shortauthors{Pooley et al.}
\slugcomment{Accepted by the {\it Astrophysical Journal}}
\begin{document}

\submitted{Accepted by the {\it Astrophysical Journal}}

\title{Chandra Observation of the Globular Cluster NGC~6440 and the
Nature of Cluster X-ray Luminosity Functions}

\author{David Pooley\altaffilmark{1},
Walter H.\ G.\ Lewin\altaffilmark{1}, 
Frank Verbunt\altaffilmark{2},
Lee Homer\altaffilmark{3},
Bruce Margon\altaffilmark{4},
Bryan M.\ Gaensler\altaffilmark{5},
Victoria M.\ Kaspi\altaffilmark{6,1},
Jon M.\ Miller\altaffilmark{1}, 
Derek W.\ Fox\altaffilmark{7}, 
Michiel van der Klis\altaffilmark{8}}

\altaffiltext{1}{Center for Space Research and Department of Physics,
Massachusetts Institute of Technology, Cambridge, MA  02139-4307;
davep@space.mit.edu, lewin@space.mit.edu, jmm@space.mit.edu} 
\altaffiltext{2}{Astronomical Institute, Utrecht University, PO Box
80 000, 3508 TA Utrecht, The Netherlands; F.W.M.Verbunt@astro.uu.nl}
\altaffiltext{3}{Astronomy Department, Box 351580, University of
Washington, Seattle, WA 98195; homer@astro.washington.edu}
\altaffiltext{4}{Space Telescope Science Institute, 3700 San Martin
Drive, Baltimore, MD 21218; margon@stsci.edu}
\altaffiltext{5}{Harvard-Smithsonian Center for Astrophysics, 60
Garden Street, Cambridge, MA 02138; bgaensler@cfa.harvard.edu}
\altaffiltext{6}{Department of Physics, Rutherford Physics Building, McGill University, 3600
University Street, Montreal, QC H3A 2T8, Canada; vkaspi@physics.mcgill.ca} 
\altaffiltext{7}{Astronomy Department, California Institute of
Technology, Mail Code 105-24, Pasadena, CA 91125;
derekfox@astro.caltech.edu}
\altaffiltext{8}{Astronomical Institute ``Anton Pannekoek,''
University of Amsterdam and Center for High-Energy Astrophysics,
Kruislaan 403, 1098 SJ Amsterdam, The Netherlands; michiel@astro.uva.nl}

\begin{abstract}
As part of our campaign to determine the nature of the various source
populations of the low-luminosity globular cluster X-ray sources, we
have obtained a \cxo\ ACIS-S3 image of the globular cluster NGC~6440.
We detect 24 sources to a limiting luminosity of
$\sim$\ee{2}{31}~\ergsec\ (0.5--2.5~keV) inside the cluster's
half-mass radius, all of which lie within $\sim$2 core radii of the
cluster center. We also find excess emission in and around the core
which could be due to unresolved point sources.  Based upon X-ray
luminosities and colors, we conclude that there are 4--5 likely
quiescent low-mass X-ray binaries and that most of the other sources
are cataclysmic variables.  We compare these results to \chandra\
results from other globular clusters and find the X-ray luminosity
functions differ among the clusters.
\end{abstract}

\keywords{globular clusters: individual (NGC~6440) --- stars: neutron
--- cataclysmic variables --- binaries: close}

\section{Introduction}
NGC~6440 is a globular cluster near the center of our Galaxy, at a
distance of $8.5\pm0.4$~kpc and reddened by $E_{B-V}=1.07$
\citep{orto94}.  It is one of the twelve globular clusters in which a
bright ($\Lx>10^{36}$~\ergsec) X-ray source has been detected so far.
The source in NGC~6440 is a transient, first detected during an
outburst which lasted more than a month, in December 1971 and January
1972 \citep{mark75,form76}.  During the outburst, the 2--11~keV
luminosity was more or less constant, at \ee{3}{37}~\ergsec.  A second
outburst was detected in August 1998. The observed luminosity was
$<10^{37}$~\ergsec\ \citep{intz99} at peak and dropped by a factor of 400 in less
than 17 days \citep{verb00}. A third outburst was recently detected in
August 2001 with the All Sky Monitor (ASM) on the {\it Rossi X-ray
Timing Explorer} and announced by the ASM team at
\url{http://xte.mit.edu}.  A \chandra\ observation showed the
luminosity to be \ee{9}{35}~\ergsec\ \citep{intz01}.  The difference
in the properties of the outbursts raises the question whether they
were indeed from the same source or different sources, and, if the
former, how outbursts of the same source can be so different.

Comparison of optical images of NGC~6440 taken before and during the
1998 outburst show one star that was brighter in $B$, but not in $R$,
during the outburst. This star, estimated at $B=22.7$ and $(B-R)_0<0$
near the end of the outburst, is a viable optical counterpart to the
X-ray transient \citep{verb00}.

Between outbursts, the core of NGC~6440 was detected by the {\it
Einstein} and {\it ROSAT} satellites \citep{hg83,jvh95}. A longer {\it ROSAT}
observation showed that the core contains (at least) two X-ray
sources \citep{verb00}.

The high incidence of bright X-ray sources in globular clusters is
likely explained by the formation of binaries containing a neutron
star via tidal capture and/or exchange encounters.  In a tidal
capture, a neutron star passing close to another star transfers enough
of its kinetic energy to the tidal bulge of the other star to keep it
bound. In an exchange encounter, a neutron star takes the place of a
binary member by expelling it (for a review of these processes, see
Hut et al.\ 1992). To first order, both processes scale with the
number of collisions in a cluster, i.e., with the product of the
volume of the core (where most close encounters occur) and the square
of the number density of stars in the core. NGC~6440 has one of the
highest collision numbers, exceeding that of 47~Tuc, according to the
best current values of cluster parameters (Table~\ref{tab:params}). In
dense clusters like NGC~6440 and 47~Tuc, the formation of cataclysmic
variables is also dominated by close encounters.

In this paper, we describe our observation of NGC~6440 with \chandra,
aimed at discovering how many and what type of X-ray sources it
contains and at finding out whether any of these corresponds to the
star that is a possible counterpart for the 1998 transient.  Similar
results from \chandra\ have been reported for the other clusters
47~Tuc \citep{grin01a}, $\omega$~Cen \citep{rutl01}, NGC~6397
\citep{grin01b}, and NGC~6752 \citep{pool01}.

\begin{figure*}[ht!]
\plotone{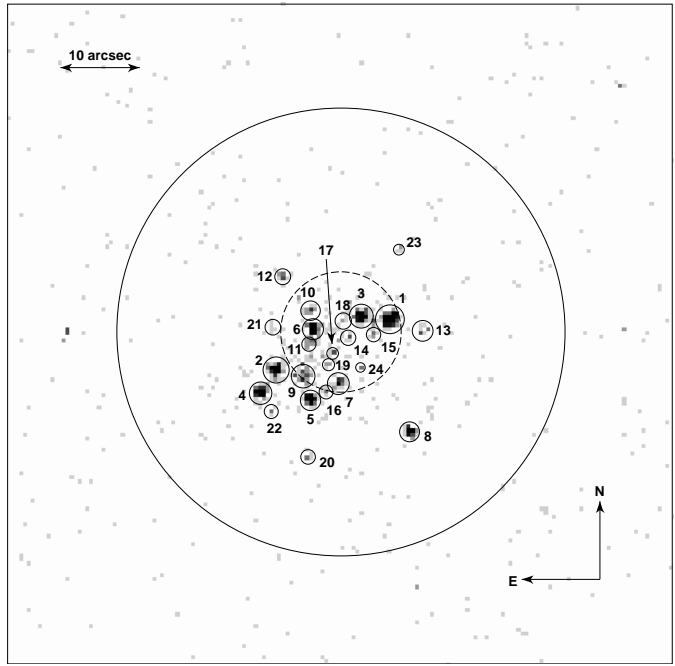}
\caption{X-ray image of the central $1\farcm4 \times 1\farcm4$ region
of NGC~6440.  The numbered solid circles indicate the source
extraction regions.  The dashed circle represents the 7\farcs58 core
radius, and the solid circle is the 28\farcs2 half-mass radius.  Each
pixel is a 0\farcs492 square.\label{fig:xrayimage}}
\end{figure*}

\section{X-ray Observations and Analysis}
NGC~6440 was observed for 23~ks on 2000~July~4 with the Advanced CCD
Imaging Spectrometer (ACIS) on the \cxo\ with the telescope aimpoint
on the back-side illuminated S3 chip. The data were taken in
timed-exposure mode with the standard integration time of 3.24~s per
frame and telemetered to the ground in faint mode.

Data reduction was performed using the CIAO~2.1 software provided by
the \chandra\ X-ray Center (\url{http://asc.harvard.edu}).  We used
the CALDB\,2.8 set of calibration files (gain
maps, quantum efficiency, quantum efficiency uniformity, effective
area).  Bad pixels were excluded.  Intervals of background
flaring were searched for, but none were found.  

Starting with the raw (level~1) event list, we processed the data
(using {\tool acis\_process\_events}) without including the pixel
randomization that is added during standard processing.  This method
slightly improves the point spread function (PSF).  We then applied
the good-time intervals supplied with the standard data
products and filtered the data to include only events with {\it ASCA}
grades of 0, 2, 3, 4, or 6 (this is the ``standard'' choice that
generally optimizes the signal-to-background ratio; see the \chandra\
Proposer's Observatory Guide available from the website for more
information).  We also excluded software-flagged cosmic ray events.
We used this filtered event list (level~2) for the subsequent
analysis.

\subsection{Source Detection} \label{sect:detection}
The wavelet-based {\tool PWDetect} tool \citep{dam97} was employed for
source detection in the 0.5--6.0~keV band.  We found 24 point sources
within the 28\farcs2 cluster half-mass radius \citep{tra93} and
another 64 on the rest of the S3 chip.  Table~\ref{tab:srcs} lists the
cluster sources.  We have numbered the sources in order of detected
counts in the 0.5--6.0~keV band.  Our detection threshold was $\geq$3
counts ($\sim$10$^{-15}$~\ergcms).  The density of sources outside the
half-mass radius implies that 0.5 sources within the half-mass
radius are not associated with the cluster.  This is in agreement
with the expected number of background sources from the
$\log{N}-\log{S}$ relationships of \citet{gia01}.  All 24 possible
cluster sources are consistent with being point sources, with the
exception of CX9, which is inconsistent at $>5\sigma$.  It is likely a
blend of multiple sources.

A $1\farcm4 \times 1\farcm4$ region centered on the core is shown in
Fig.~\ref{fig:xrayimage}.  The extraction regions for each source are
shown, as are circles indicating the core radius (7\farcs58; Trager et
al.\ 1993) and half-mass radius.  We note that there is excess
emission in and around the core.  The residual count rate (i.e., the
total rate minus the rate due to all {\tool PWDetect} sources) in the
0.5--6~keV band inside a 15\arcsec\ radius region in the center of the
cluster is $(5.3\pm0.4)\times10^{-4}$~cts~pixel$^{-1}$~ks$^{-1}$,
which is a factor of $5.1\pm 0.6$ higher than the rate from
source-free regions outside the cluster. This emission is probably due
to many unresolved point sources.

\subsection{Astrometry} \label{sec:astrometry}
We have reanalyzed 1.5~GHz data from the Very Large Array (VLA;
Napier, Thompson, \& Ekers 1983) taken on 1990 April 20 at in A
configuration and 1990 July 8 in BnA configuration.  We detect three
point sources in the region covered by the ACIS-S3 chip
(Table~\ref{tab:radiosrcs}); their positions are in agreement with the
sources found by \citet{knapp96} and \citet{fg00}.  The two sources
outside the cluster both lie near {\tool PWDetect} sources.  The third
source, PSR~B1745$-$20, falls in between two X-ray sources (CX7 and
CX19) in the core of NGC~6440.

Using the two sources outside the cluster to align the frames, we find
that a small shift of (0\farcs3, $-$0\farcs1) brings the X-ray and
radio positions into agreement for both sources.  However,
PSR~B1745$-$20 is still inconsistent with either X-ray source and is thus
undetected in this \chandra\ observation.  This is not surprising as
its spin-down luminosity is \ee{6.6}{32}~\ergsec\ \citep{taylor93}.  This implies an
X-ray luminosity $\Lx \approx \ee{7}{29}$~\ergsec\ \citep{vb96}, which is well
below our detection threshold of $\sim$\ee{2}{31}~\ergsec.



\begin{figure*}
\plotone{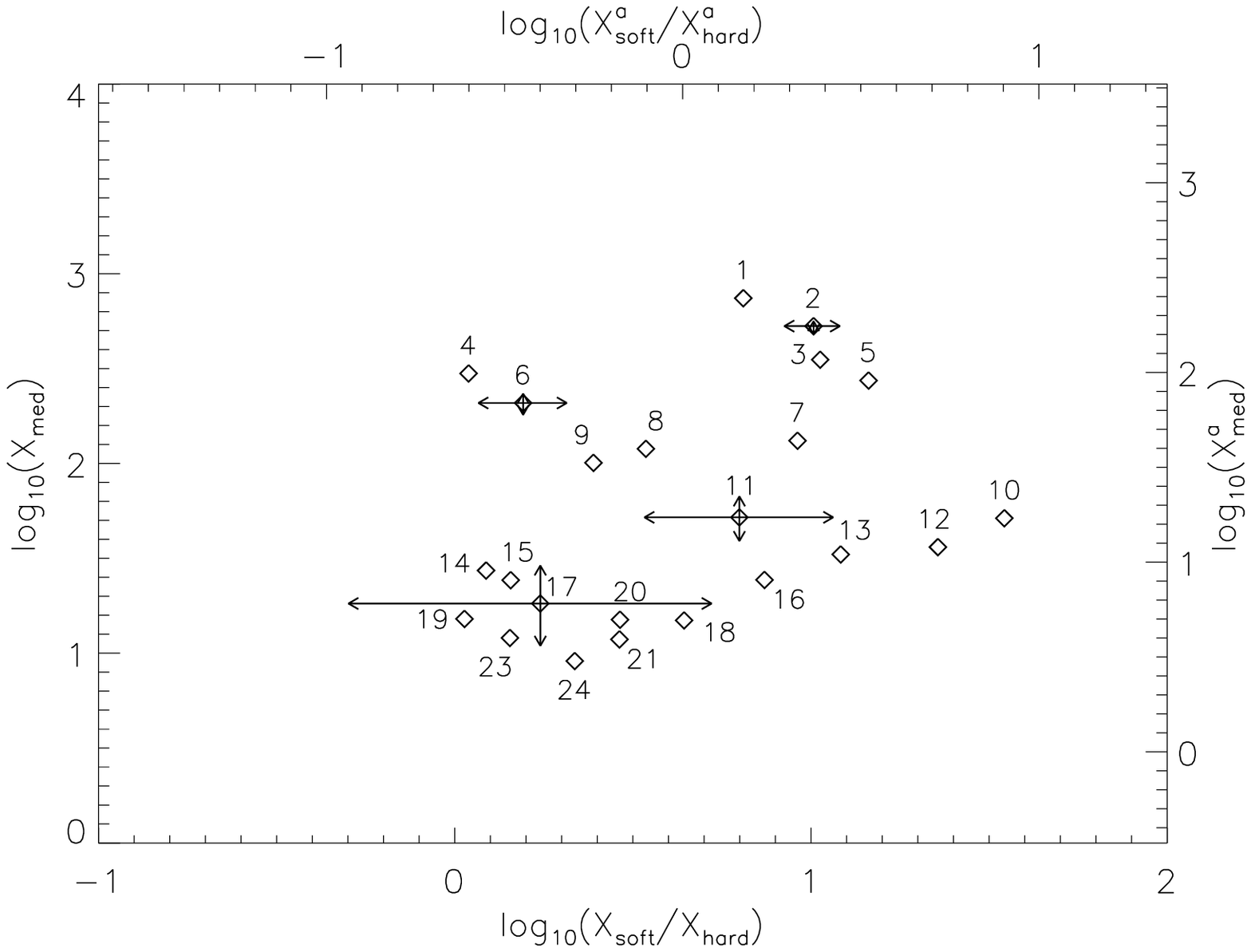}
\caption{X-ray color-magnitude diagram. The X-ray color is defined as
the logarithm of the ratio of \xsoft\ (0.5--1.5~keV) corrected counts
to \xhard\ (1.5--6.0~keV) corrected counts, and the magnitude is the
logarithm of \xmed\ (0.5--4.5~keV) corrected counts.  For the sake of
clarity, only a few error bars are shown; they represent the 1$\sigma$
error estimates given by \citet{gehr86}.  Our correction for
photoelectric absorption has the effect of uniformly shifting the data
$+$0.48 units on the left axis and $+$0.64 units on the bottom
axis. The top and right axes provide the absorbed color and magnitude
scales ($^{\rm a}$), i.e., the observed colors and magnitudes
uncorrected for absorption.  CX22 is not shown since it has 0 counts
in \xsoft.  \label{fig:xraycmd}}
\end{figure*}

\subsection{Count Rates} \label{sec:cts}
We extracted source counts in the following bands: 0.5--1.5~keV
(\xsoft), 0.5--4.5~keV (\xmed), and 1.5--6.0~keV (\xhard).  The
detected count rate was corrected for background, exposure variations,
and foreground photoelectric absorption.  We make these corrections in
order to produce an X-ray color-magnitude diagram (CMD) that can be
meaningfully compared to the X-ray CMDs that have resulted from
\chandra\ observations of other globular clusters.  In addition,
however, attention must be given to differences in detector responses
and, of course, exposure times.

The background count rate in each band was estimated from an annulus
around the core sources. The inner radius was 31~pixels, and the outer
radius was 64~pixels.  No detected sources were present within the annulus.
The density of background counts in each band (for 23~ksec) was found
to be 0.006 counts per pixel (\xsoft), 0.011 counts per pixel (\xmed),
and 0.007 counts per pixel (\xhard).  The background count rate in the
core may be somewhat higher, but even factors of a few greater than
this estimate have negligible effects on our analysis.

To account for the $\sim$1\% variations in exposure among the sources,
we applied multiplicative corrections based on the ratio of a source's
average effective area in each of the three bands to the average
effective area in the same band of CX12, which had the highest average
exposure.  The individual effective area curves for the sources were
made using the CIAO tool {\tool mkarf}.  The average effective area of
CX12 in each of the bands was 580~\cmsq\ (\xsoft), 455~\cmsq\ (\xmed),
and 368~\cmsq\ (\xhard).

While the previous corrections were relatively minor (at the few
percent level or less), the correction for photoelectric absorption is
rather large for NGC~6440.  The conversion of optical reddening to
column density \citep{pred95} gives a value of
$N_H=\ee{5.9}{21}~\pcmsq$.  We investigated the effects of such an
absorption on three characteristic spectra: a 3~keV thermal
bremsstrahlung, a 0.3~keV blackbody plus power law with photon index
of $\Gamma=2$, and a power law with a photon index of $\Gamma=2$.  The
effects were most dramatically seen in the \xsoft\ band, where the
absorbed count rate was a factor of 4.8--6.1 lower than the unabsorbed
one (depending on the spectrum).  Averaging the results of each
spectrum in each band, we use the following correction factors: 5.43
(\xsoft), 3.06 (\xmed), and 1.24 (\xhard).  Table~\ref{tab:srcs} lists
both the observed and fully corrected counts in each band.  The effect
of the absorption correction on the X-ray color-magnitude diagram
(Fig.~\ref{fig:xraycmd}) is a uniform shift of all points 0.48 units
on the left axis and 0.64 units on the bottom axis.  The top and right
axes give the X-ray color and magnitude without this shift (they do,
however, include the small corrections for background subtraction and
exposure variations).

\subsection{Spectral Fitting}
We used the CIAO tool {\tool dmextract} to extract spectra of sources
CX1--CX11 in the 0.3--8~keV range.  Because even the brightest source
(CX1) had relatively few counts, we chose to fit the unbinned spectra
using the $C$-statistic in \xspec\ \citep{arn96}.  This statistic is
insensitive to the number of counts per bin and is thus a good choice
when fitting low-count data. It should be noted, however, that the
$C$-statistic is only applicable when the background is negligible,
which is the case with these data. To determine the goodness of the
fits, we ran Monte Carlo simulations of each best-fit spectrum for
each source and compared with the observed spectra.  For a good fit,
the simulations should give lower values of the fit statistic than the
data give about 50\% of the time.

Three different models were fit: thermal bremsstrahlung (TB),
blackbody plus power-law (BB+PL), and power law (PL).  We fixed \nh\
to the value from optical reddening.  As expected for such low-count
spectra, very few fits could be formally ruled out.  We estimated the
unabsorbed source luminosities for CX1--CX11 by averaging the results
from the three best-fit models for each source.  The spread in \Lx\ of
the three models was $\sim$20\% for each source.  Fitting a
linear relation to these luminosities versus detected \xmed\ counts, we
have estimated the unabsorbed luminosities for sources CX12--CX24
based on their \xmed\ counts.  These are listed in
Table~\ref{tab:srcs}.  

\begin{figure*}
\plotone{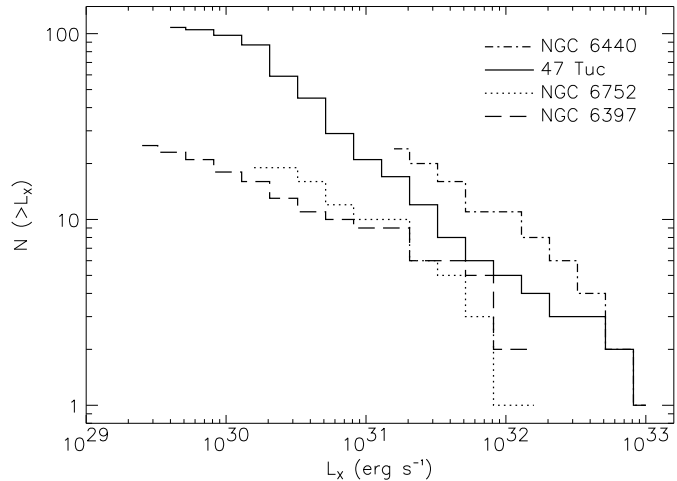}
\caption{Histogram of the number of \chandra\ sources detected above a
certain \Lx\ ($N(>\!\!\Lx)$)
versus \Lx\ [0.5--2.5~keV] for four globular clusters. \label{fig:logNlogS}}
\end{figure*}

Fitting a hydrogen atmosphere spectral model to the securely
identified quiescent neutron star CX1, \citet{intz01} find an X-ray
luminosity of \ee{1.7}{33}~\ergsec\ in the 0.5--7~keV range.  In the
0.5--2.5~keV band, the hydrogen atmosphere model gives $\Lx =
\ee{1.5}{33}~\ergsec$, which is $\sim$50\% more than our estimate.  However,
the results are consistent with each other within errors.

\section{Results and Discussion} \label{sec:res}
As shown in Fig.~\ref{fig:xrayimage}, the sources detected by
\chandra\ are concentrated towards the center of NGC~6440, and all lie
well within the half-mass radius of the cluster. This and the estimate
that we expect only $\sim$0.5 background sources in this region
suggests that all sources are cluster members.  Because of the large
distance and high reddening to NGC~6440, there is little suitably deep
optical imaging available.  In the absence of optical identifications
-- with one notable exception discussed below -- we try to classify
the sources on the basis of their X-ray properties.

In the Galactic disk, cataclysmic variables have X-ray luminosities up to
$\sim10^{32}$~\ergsec\ in the 0.5--2.5~keV range (see e.g.\ Fig.~8 in
Verbunt et al.\ 1997; the single outlier in that figure is the DQ~Her
type system V1223~Sgr at $\Lx\simeq 4\times 10^{32}$~\ergsec).  Low
mass X-ray binaries (LMXBs) in the Galactic disk reach higher
luminosities, even in quiescence, and have X-ray spectra that are soft
(blackbody color temperatures of $\sim$0.3~keV).  The faintest known
quiescent LMXB (qLMXB) with a neutron star has $\Lx\simeq 7\times
10^{31}$~\ergsec\ (Cen~X-4; Campana et al.\ 1997), but binaries with
a black hole may reach lower X-ray luminosities (e.g., A0620$-$00
dropped to $\sim$ 10$^{30}$~\ergsec, McClintock, Horne, \& Remillard
1995).  RS~CVn systems in the Galactic disk reach luminosities in the
range of the sources detected with \chandra\ in NGC~6440. However, the
brightest \chandra\ sources optically identified with RS~CVn systems
(in the wide definition which includes BY~Dra binaries) in globular
clusters have $L_{\rm x}\simeq 3, 2, 0.3\times 10^{31}$~\ergsec\ in
47~Tuc, NGC~6397, and NGC~6752, respectively
\citep{grin01a,grin01b,pool01}.

Bringing this information to bear on the sources in NGC~6440, we
conclude that CX1, CX2, CX3, and CX5 are qLMXBs, that CX4, CX6, and
CX7 could be either qLMXBs or cataclysmic variables, and that most
other sources that we have detected are cataclysmic variables. This is
further clarified by comparing Fig.~\ref{fig:xraycmd} to the X-ray
CMDs of 47~Tuc, NGC~6397, and NGC~6752 \citep{grin01a,grin01b,pool01}.
The X-ray colors of CX1, CX2, CX3, CX5, and CX7 are all $\sim$1, as is
the X-ray color of the probable qLMXB in NGC~6397 (its listed X-ray
color of $\sim$2.6 includes multiplication by a factor of 2.5 which
Grindlay et al.\ use to define their X-ray colors and magnitudes).
The two probable qLMXBs in 47~Tuc have X-ray colors of $\sim$0.6
(again, without the factor of 2.5).  In contrast, the brightest CVs in
47~Tuc, NGC~6397, and NGC~6752 all have X-ray colors of $\sim$0, and
the vast majority of identified CVs in these three clusters have X-ray
colors in the range $-$0.5 to 0.5, suggesting that CX7 is a qLMXB and
that CX4 and CX6 are CVs.  Because of their relatively soft X-ray
colors, CX10, CX12, and CX13 might be quiescent transients with a
black hole, but they are more likely cataclysmic variables.

Some of the faintest sources that we detected may be RS~CVn systems or
millisecond pulsars, which have X-ray luminosities in this range (see
Fig.~8 in Verbunt et al.\ 1997).

CX1 coincides within the positional error with the optical variable
V2, the suggested counterpart for the transient in NGC~6440 that went
into a short faint outburst in August 1998 \citep{verb00}.  Since CX1
shows the X-ray characteristics of a transient in quiescence, and V2
the optical characteristics (magnitude and color) of such a source in
outburst, we consider the identification secure.  The transient showed
X-ray bursts \citep{intz99,intz01}, indicating that the accretor in
CX1 is a neutron star.  This association of CX1 and V2 confirms the
suggested optical identification \citep{verb00} of this first qLMXB to
be identified in outburst.

Based on the secure CX1/V2 identification, we shift the optical frame
of \citet{verb00} into the \chandra\ frame of this observation.  After
this small ($\sim$0\farcs37) shift, we find that CX24 is coincident to
$\sim$0\farcs1 with star V0 of \citet{verb00}, a suggested Mira type
variable.  Since Mira's are cool giants, they are not expected to be
X-ray sources.  Remarkably, \citet{bell98} tentatively identify an
X-ray source in the old open cluster M67 with an X-ray luminosity very
similar to that of CX24 with a star near the tip of the giant branch
(X19/S364 in their Table~3). We therefore do not discard offhand the
possible identification of CX24 with V0.

\begin{figure*}
\plotone{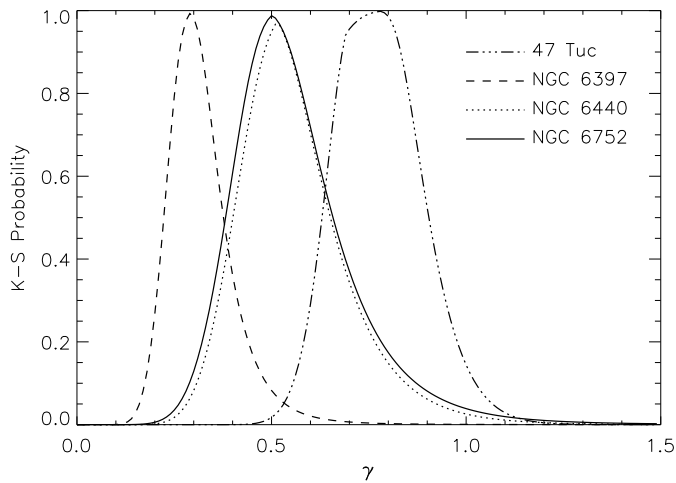}
\caption{The Kolmogorov-Smirnov probability, as a function of
$\gamma$, that the $z^j = (L^j_i/L_i)^{-\gamma}$ are uniformly
distributed. $L_i$ is the limiting \Lx\ to which cluster $i$ has been
observed, $L_i^j$ is the \Lx\ of source $j$ in cluster $i$, and
$\gamma$ is the index of the luminosity function
$dN\propto L_{\rm x}^{-\gamma}\, d\log{\Lx}$. \label{fig:ks}}
\end{figure*}

The number of close encounters in a globular cluster scales with
${\rho_{\rm o}}^2{r_{\rm c}}^3$, where $\rho_{\rm o}$ is the central
density of the cluster and $r_{\rm c}$ its core radius.  The number of
low-mass X-ray binaries, and the number of cataclysmic variables are
both thought to scale with the number of close encounters in clusters
with dense cores like 47~Tuc and NGC~6440. This implies that the
numbers of these sources in NGC~6440 should be $\sim$1.7 times that in
47~Tuc.  Two of the 20 X-ray sources in 47~Tuc with $L_{\rm x}>
10^{31}$ erg~s$^{-1}$ are probably qLMXBs, 17 of the other sources are
cataclysmic variables, and one is a BY~Dra system.  The number of 4 or
5 qLMXBs in NGC~6440 is compatible with this prediction, but the
number of 19 or 18 cataclysmic variables in the same luminosity range
is somewhat lower than predicted.  However, the excess flux above
background (see \S\ref{sect:detection}) in and around the core
corresponds to a luminosity of $\sim$\ee{5}{32}~\ergsec, and has a
corrected X-ray color of 0.5.  This emission could be due to
unresolved point sources.

At the same time, destruction of close binaries by multiple close
encounters depresses the numbers of X-ray sources in the densest
clusters, according to theory.  From a study of the \rosat\ data,
\citet{john96} found that the number of faint sources actually scales
with ${\rho_{\rm o}}^{1.5}{r_{\rm c}}^3$, not ${\rho_{\rm o}}^2{r_{\rm
c}}^3$, which appears to confirm this prediction.  As \chandra\ data
of more clusters become available, it will be interesting to verify
this in more detail; the possible relatively low number of X-ray
sources in NGC~6440 may be a first hint.  However, given the
uncertainty in the collision numbers and the possibility of many
unresolved sources, we refrain from drawing premature conclusions.

From the \rosat\ data, \citet{john96} also derive a power-law
luminosity function $dN\propto L_{\rm x}^{-\gamma}\, d\ln{\Lx}$ with
$\gamma=0.5$.  The \chandra\ observations have resolved several
\rosat\ sources into multiple sources and have also detected sources
with lower luminosities. As a result, the number of faint sources is
already larger in the few clusters observed with \chandra\ than in the
complete \rosat\ data set described by \citet{verb01}.  We plot the
cumulative luminosity functions of four clusters observed with
\chandra\ in Fig.~\ref{fig:logNlogS}.  Following the statistical
method of \citet{john96}, we derive the best-fit luminosity function
by forming the quantities $z^j = (L^j_i/L_i)^{-\gamma}$ and finding
the $\gamma$ that most uniformly distributes the $z^j$ along the
segment [0,1].  Here, $L_i$ is the limiting \Lx\ to which cluster $i$
has been searched, and $L_i^j$ is the \Lx\ of source $j$ in cluster
$i$.  In order to reduce errors in the analysis due to incompleteness
of the sample at the low end of the luminosity range, we have taken
each $L_i$ to be $\sim$3 times the \Lx\ of the faintest observed
source in each cluster and have restricted the data accordingly.  We
examined the uniformity of the $z^j$ distribution for a range of
$\gamma$ through Kolmogorov-Smirnov tests.  The results of these tests
are shown in Fig.~\ref{fig:ks}.  The best-fit $\gamma$ and the
corresponding probabilities of the fits are given in
Table~\ref{tab:logNlogS}.

To further investigate the luminosity functions, we collect in
Table~\ref{tab:logNlogS} the numbers of faint sources detected with
Chandra in four clusters, as a function of their luminosity. The
surveyed area in each cluster is well within the half-mass radius, and
most sources are expected to be associated with the clusters.
According to the Johnston-Verbunt result, the number of sources in a
logarithmic bin, corresponding to one decade of luminosities,
increases by a factor $\sqrt{10}$ between adjacent bins.
Table~\ref{tab:logNlogS} shows that the current data tend to a
somewhat slower increase towards lower luminosities in three of the
four clusters. This trend is especially significant in NGC~6397, in
marked contrast to the cluster 47~Tuc in which the increase towards
lower luminosities is slightly faster than a factor 10 over two bins.
In Table~\ref{tab:logNlogS} we note an anticorrelation between the
slope of the luminosity function and the central cluster density. We
investigate two possible explanations. First, in the densest clusters,
mass segregation may enhance the neutron-star capture rate (e.g.,
Verbunt \& Meylan 1988) and thus enhance the more luminous population
of qLMXBs. The numbers of such sources, however, are too small in the
clusters observed to strongly affect the luminosity function. Indeed,
the numbers in Table~\ref{tab:logNlogS} show that the difference
between NGC~6397 and 47~Tuc, the most and least dense clusters, lies
mainly in the low-luminosity sources, i.e., presumably the RS~CVn and
CV systems. This difference may be due to the destruction of RS~CVns
and CVs (or CV progenitors) in the densest clusters by encounters
with neutron stars. Observations of a larger number of clusters may
help in confirming the variation in luminosity function among globular
clusters and in explaining it as a function of the initial cluster
properties.  \chandra\ observations combined with simulations with the
GRAPE-6 \citep{ma01} could be instrumental in understanding the dynamical
evolution of globular clusters.

\acknowledgments
DP acknowledges that this material is based upon work partly supported under
a National Science Foundation Graduate Fellowship.  
WHGL gratefully acknowledges support from NASA.
LH acknowledges the support of NASA through LTSA grant NAG5-7932.
BMG acknowledges the support of a Clay Fellowship awarded by the
Harvard-Smithsonian Center for Astrophysics.
VMK is a Canada Research Chair and acknowledges support from LTSA
grant NAG5-8063, NSERC Rgpin 228738-00, and a Sloan Fellowship.
MvdK acknowledges financial support from the Netherlands Organization for
Scientific Research (NWO).
The National Radio Astronomy Observatory is a facility of the National
Science Foundation operated under cooperative agreement by Associated
Universities, Inc.

\clearpage

\begin{deluxetable}{lcll}
\tablewidth{0pt}
\tablecaption{Globular Cluster Physical Parameters. \label{tab:params}}
\tablecolumns{4}
\tablehead{
\colhead{Cluster}& \colhead{$\rho_0$ ($L_\sun$ pc$^{-3}$)}& \colhead{$r_c$}& \colhead{$M_V$}
}
\startdata
47 Tuc &   4.77 & 23\arcsec & -9.42 \\
NGC 6397 & 5.68 & 3\arcsec &  -6.63 \\
NGC 6440 & 5.28 & 7\farcs6 & -8.75 \\
NGC 6752 & 4.91 & 10\farcs5 & -7.73 \\
\enddata
\tablecomments{Central density, core radius, and integrated visual
magnitude for four globular clusters.  Values are from the catalog of
\citet{har96} except for the core radii of 47~Tuc \citep{how00},
NGC~6440 \citep{tra93}, and NGC~6752 \citep{tra93}.}
\end{deluxetable}

\begin{deluxetable}{lllllll}
\tablewidth{0pt}
\tablecaption{NGC~6440 X-ray Sources. \label{tab:srcs}}
\tablecolumns{7}
\tablehead{
\colhead{Src\tablenotemark{a}}& \colhead{RA (J2000)\tablenotemark{b}}& \colhead{Dec (J2000)\tablenotemark{b}}& \multicolumn{3}{c}{Detected
Counts/Corrected Counts\tablenotemark{c}} & \colhead{\Lx\ (\ergsec)\tablenotemark{d}} \\
\colhead{} & \colhead{} & \colhead{} &
\colhead{\xsoft} & \colhead{\xmed} & \colhead{\xhard} & \colhead{[0.5--2.5~keV]}
}
\startdata
CX1&   17 48 52.163& $-$20 21 32.40& 147/800& 242/740&  100/120& \ee{9.9}{32} ($\pm$13\%)\\
CX2&   17 48 53.181& $-$20 21 38.78& 120/660& 172/530&  52/65  & \ee{8.4}{32} ($\pm$19\%)\\
CX3&   17 48 52.419& $-$20 21 32.00& 82/450&  115/350&  34/42  & \ee{5.3}{32} ($\pm$14\%)\\
CX4&   17 48 53.320& $-$20 21 41.70& 22/120&  97/300 &  88/110 & \ee{2.3}{32} ($\pm$14\%)\\
CX5&   17 48 52.874& $-$20 21 42.61& 69/380&  89/270 &  21/26  & \ee{5.0}{32} ($\pm$15\%)\\
CX6&   17 48 52.852& $-$20 21 33.64& 21/110&  68/210 &  59/73  & \ee{2.4}{32} ($\pm$30\%)\\
CX7&   17 48 52.625& $-$20 21 40.50& 29/160&  43/130 &  14/17  & \ee{2.0}{32} ($\pm$21\%)\\
CX8&   17 48 51.987& $-$20 21 46.57& 18/98&   39/120 &  23/29  & \ee{1.3}{32} ($\pm$20\%)\\
CX9&   17 48 52.939& $-$20 21 39.56& 14/76&   33/100 &  25/31  & \ee{1.0}{32} ($\pm$20\%)\\
CX10&  17 48 52.874& $-$20 21 31.31& 15/81&   17/52 &   2/2    & \ee{1.7}{32} ($\pm$25\%)\\
CX11&  17 48 52.888& $-$20 21 35.53& 10/54&   17/52 &   7/9    & \ee{1.3}{32} ($\pm$27\%)\\
CX12&  17 48 53.123& $-$20 21 27.03& 10/54&   12/36 &   2/2    & \ee{5.5}{31} ($\pm$39\%)\\
CX13&  17 48 51.870& $-$20 21 33.85& 8/43&    11/33 &   3/4    & \ee{5.1}{31} ($\pm$40\%)\\
CX14&  17 48 52.537& $-$20 21 34.73& 2/11&    9/27  &   7/9    & \ee{4.3}{31} ($\pm$45\%)\\
CX15&  17 48 52.310& $-$20 21 34.33& 2/11&    8/24  &   6/7    & \ee{3.9}{31} ($\pm$47\%)\\
CX16&  17 48 52.734& $-$20 21 41.56& 5/27&    8/24  &   3/4    & \ee{3.9}{31} ($\pm$47\%)\\
CX17&  17 48 52.676& $-$20 21 36.70& 2/11&    6/18  &   5/6    & \ee{3.0}{31} ($\pm$54\%)\\
CX18&  17 48 52.581& $-$20 21 32.62& 3/16&    5/15  &   3/4    & \ee{2.6}{31} ($\pm$60\%)\\
CX19&  17 48 52.712& $-$20 21 38.24& 1/5&     5/15  &   4/5    & \ee{2.6}{31} ($\pm$60\%)\\
CX20&  17 48 52.896& $-$20 21 49.85& 2/11&    5/15  &   3/4    & \ee{2.6}{31} ($\pm$60\%)\\
CX21&  17 48 53.210& $-$20 21 33.39& 2/11&    4/12  &   3/4    & \ee{2.1}{31} ($\pm$67\%)\\
CX22&  17 48 53.225& $-$20 21 44.01& 0/--&    4/12  &   5/6    & \ee{2.1}{31} ($\pm$67\%)\\
CX23&  17 48 52.083& $-$20 21 23.63& 1/5&     4/12  &   3/4    & \ee{2.1}{31} ($\pm$67\%)\\
CX24&  17 48 52.427& $-$20 21 38.46& 1/5&     3/9  &    2/2    & \ee{1.8}{31} ($\pm$78\%)\\
\enddata
\tablenotetext{a}{Sources are numbered according to their detected
counts in the 0.5--6~keV band.}
\tablenotetext{b}{Nominal \chandra\ positions from {\tool PWDetect}.
The pointing uncertainty is $\sim$0.6\arcsec.}
\tablenotetext{c}{Corrections are described in \S\ref{sec:cts}.
X-ray bands are 0.5--1.5~keV (\xsoft), 0.5--4.5~keV (\xmed), and
1.5--6.0~keV (\xhard).}
\tablenotetext{d}{For sources CX1--CX11, \Lx\ comes from an average of
the unabsorbed luminosities of the best-fit models for each source.
A linear relation between \Lx\ and \xmed\ counts for these sources was
derived and used to estimate \Lx\ for sources CX12--CX24 based upon
their \xmed\ counts.  Uncertainties for CX1--CX11 come mainly from the
spread in \Lx\ from the three best-fit models; the distance and
reddening uncertainties are negligible in comparison.  For CX12--CX24,
the uncertainties are a combination of the \Lx-\xmed\ fit uncertainty
and the Poisson uncertainty for each source.}
\end{deluxetable}

\begin{deluxetable}{llllll}
\tablewidth{0pt}
\tablecaption{Radio Sources in and around NGC~6440. \label{tab:radiosrcs}}
\tablecolumns{6}
\tablehead{
\colhead{Src\tablenotemark{a}} & \multicolumn{2}{c}{VLA Position} &
\multicolumn{2}{c}{\chandra\ Position} & \colhead{Offset}\\
\colhead{}& \colhead{RA (J2000)\tablenotemark{b}} & \colhead{Dec
(J2000)\tablenotemark{c}} & \colhead{RA (J2000)\tablenotemark{b}} &
\colhead{Dec (J2000)\tablenotemark{c}} & \colhead{After
Shift\tablenotemark{d}}
}
\startdata
A&$46.345\pm0.002$& 21 $36.12\pm0.02$&
$46.333\pm0.021$& 21 $36.33\pm0.32$&
($-0\farcs11\pm0\farcs30$, $0\farcs11\pm0\farcs32$)\\
B&$47.985\pm0.001$& 19 $59.28\pm0.01$& $47.915\pm0.053$&
19 $58.34\pm0.80$& ($0\farcs70\pm0\farcs74$, $0\farcs84\pm0\farcs80$)\\
C&$52.70\pm0.02$& 21 $39.29\pm0.15$& 
\parbox{6.3em}{$52.625\pm0.014$\\ $52.712\pm0.019$}& 
\parbox{6.7em}{21 $40.50\pm0.22$\\ 21 $38.24\pm0.29$}&
\parbox{13em}{($0\farcs70\pm0\farcs34$, $1\farcs12\pm0\farcs27$)\\ ($-0\farcs07\pm0\farcs39$, $-1\farcs22\pm0\farcs33$)}\\
\enddata
\tablenotetext{a}{Src A corresponds to source 2 from \citet{knapp96}
and src B to their source 3.  Src C is PSR~B1745$-$20; the first
\chandra\ position listed for src C is CX7, and the second is CX19,
neither of which is consistent with the pulsar's position (i.e., the
offset is inconsistent with zero within errors).}
\tablenotetext{b}{Seconds of time added to 17$^{\rm h}$48$^{\rm m}$.}
\tablenotetext{c}{Arcminutes and arcseconds subtracted from $-$20\degr.}
\tablenotetext{d}{The shift is described in \S\ref{sec:astrometry}.}
\end{deluxetable}

\begin{deluxetable}{llllllllll}
\tablewidth{0pt}
\tablecaption{X-ray Luminosity Functions for Four Globular Clusters
Observed with \chandra. \label{tab:logNlogS}}
\tablehead{
\colhead{Cluster} & \colhead{$r_{\rm s}$} &
\colhead{$L_{\rm max}$} &
\colhead{A} & \colhead{B} & \colhead{C} &  \colhead{$\rho_0$}& \colhead{$\gamma$} &
\colhead{K-S Prob.}& \colhead{Ref.}
}
\startdata
47~Tuc&    86\arcsec & 33.0 & 5 & 15 & 74 &  4.77& $0.78\err{0.16}{0.17}$& 99.8\% &1\\
NGC~6397& 119\arcsec & 32.3 & 5 &  8 & 12 &  5.68& $0.29\err{0.11}{0.08}$& 99.3\% &2\\
NGC~6440&  16\arcsec & 33.0 & 9 & 14 &    &  5.28& $0.51\err{0.19}{0.14}$& 96.7\% &3\\
NGC~6752& 109\arcsec & 32.3 & 7 & 12 &    &  4.91& $0.50\err{0.21}{0.15}$& 98.8\% &4\\
\enddata
\tablecomments{The source with the largest distance to the cluster
center provides a measure of the surveyed area.  For each cluster, we
give this distance $r_{\rm s}$, the (logarithm of the) X-ray
luminosity $L_{\rm max}$ (in \ergsec) of the brightest source
detected, and the numbers of sources with (A) $0.1<\Lx/L_{\rm
max}\leq1.0$, with (B) $0.01<\Lx/L_{\rm max}\leq0.1$, and (where
possible) with (C) $0.001<\Lx/L_{\rm max}\leq0.01$.  We list the
logarithm of the central luminosity density $\rho_0$ (in
$L_\sun$~pc$^{-3}$) as given by \citet{har96}.  We also give the index
of the best-fit luminosity function ($dN/d\Lx\propto
\Lx^{-(\gamma+1)}$) and the Kolmogorov-Smirnov probability of the fit
for the data in Fig.~\ref{fig:logNlogS}.}  \tablerefs{1.\
\citet{grin01a}, 2.\ \citet{grin01b}, 3.\ (this work), 4.\
\citet{pool01}}
\end{deluxetable}

\end{document}